# Using Structure-Behavior Coalescence Method for Systems Definition 2.0

William S. Chao

*Abstract*—Systems definition is an artifact created by humans to describe what a system is. A system has been defined, by systems definition 1.0, hopefully to be an integrated whole, embodied in its components, their interrelationships with each other and the environment, and the principles and guidelines governing its design and evolution. This systems definition 1.0 defining the system possesses one cardinal deficiency. The deficiency comes from that it does not describe the integration of systems structure and systems behavior. Structure-behavior coalescence (SBC) architecture provides an elegant way to integrate the structure and behavior of a system. A system is therefore redefined, by systems definition 2.0, truly to be an integrated whole, using the SBC architecture, embodied in its assembled components, their interactions with each other and the environment, and the principles and guidelines governing its design and evolution. Since systems definition 2.0 describes the integration of systems structure and systems behavior, definitely it is able to form an integrated whole of a system. In this situation, systems definition 2.0 is fully capable of defining a system.

## I. INTRODUCTION

The word "system" originates from the Greek term, systēma, meaning "composition" or "whole". The notion of systems has been so extensively used in all kinds of scientific studies such as systems theory, systems analysis and design, systems architecting, systems engineering, systems ecology, systems thinking, systems bible, systems medicine, systems modeling, systems biology, systems requirement, system dynamics, etc. All things that strike us as something independent are essentially parts of a system. We usually call the parts of a system its components. Components are sometimes labeled as objects, parts, entities and building blocks. The need for systems definition arises because any real-life system is inherently complicated. It is impossible to comprehend fully the intricate interrelationships of any system of the real world with its environment, or to describe all its components and each of its details. Systems definition is an "artifact" created by humans to describe what the system is [1]. Without systems definition, everyone has his own argument about a system, never be able to reach a consensus.

A system is exceptionally complex that it includes multiple views such as strategy, concept, analysis, design, implementation, structure, behavior and input/output, and so on. The systems definition describes and represents the system multiple views possibly using two different approaches. The first one is the non-architectural approach and the second one is the architectural approach [2-3]. The non-architectural approach, also known as the model multiplicity approach [4-5], respectively picks a model for each view. The strategy view has the strategy model, the concept view has the concept model, the analysis view has the analysis model, the design view has the design model, the implementation view has the implementation model, the structure view has the structure model, the behavior view has the behavior model, and the input/output data view has the input/output data model. These multiple models, are heterogeneous and not related to each other, and thus become the primary cause of model multiplicity problems. The architectural approach, also known as the model singularity approach, instead of picking many different models, will use only one single model. The strategy, concept, analysis, design, implementation, structure, behavior, and input/output data views are all integrated in this multiple views coalescence (MVC) model of systems architecture.

## II. LITERATURE REVIEW

Systems architecture has been defined as a coalescence model of multiple views [2]. Multiple views coalescence uses only a single coalescence model. Strategy, concept, analysis, design, implementation, structure, behavior and input/output data views are all integrated in this MVC architecture. Generally, MVC architecture is synonymous with the systems architecture. In other words, multiple views coalescence sets a path to achieve the systems architecture. In the MVC architecture, multiple views must be attached to or built on the systems structure. In other words, multiple views shall not exist alone; they must be loaded on the systems structure just like a cargo is loaded on a ship. By integrating the systems structure and systems behavior, we obtain unifying structure and behavior within the system structure-behavior coalescence (SBC) [6-7] method has never been used in any systems definition for systems development except the SBC architecture. There are many advantages to use the SBC approach to integrate the systems structure and systems behavior. SBC architecture uses a single model. Systems structures and systems behaviors are integrated in this SBC architecture. Since systems structures and systems behaviors are tightly integrated, we sometimes claim that the core theme of SBC architecture is: "Systems Architecture = Systems Structure + Systems Behavior".

So far, systems behaviors are separated from systems structures in most cases. For example, the well-known structured systems analysis and design (SSA&D) approach uses structure charts (SC) to represent the systems structure and data flow diagrams (DFD) to represent the systems behavior [8]. SC and DFD are two different models and are so separated like that there is "Pacific Ocean" between them. Since structure and behavior views are the two most prominent ones among multiple views, integrating the structure and behavior views is clearly the best way to integrate multiple views of the system. In other words, structure-behavior coalescence facilitates multiple views coalescence. Multiple views coalescence sets a path to achieve the desired systems architecture with the most efficient approach. Structure-behavior coalescence facilitates multiple views coalescence. Combining the above two declarations, we conclude that structure-behavior coalescence sets a path to achieve the systems architecture. In this case, SBC architecture is also synonymous with the systems architecture. SBC architecture strongly demands that the structure and behavior views must be coalesced and integrated. This never happens in other architectural approaches such as Zachman Framework, The Open Group Architecture Framework (TOGAF), Department of Defense Architecture Framework (DoDAF) and Unified Modeling Language (UML). Zachman Framework does not offer any mechanism to integrate the structure and behavior views [9]. TOGAF, DoDAF and UML do not, either [10-12]. In the SBC architecture, a systems behavior must be attached to or built on a systems structure. In other words, a systems behavior can not exist alone; it must be loaded on a systems structure just like a cargo is loaded on a ship. There will be no systems behavior if there is no systems structure. A stand-alone systems behavior is not meaningful.

III. SBC METHOD FOR SYSTEMS DEFINITION 2.0

In this section, we will elaborate in detail the SBC architecture for systems definition 2.0. The SBC architecture for systems definition 2.0 is also called SBC process algebra (SBC-PA). In SBC-PA, each state is regarded as a process.

*A. Channel-Based Value-Passing Interactions*

A channel is a mechanism for agent communication via message passing. A message may be sent over a channel, and another agent is able to receive messages sent over a channel it has a reference to. Each channel defines a set of parameters that describes the arguments passed in with the request, or passed back out once a request has been handled. The signature for a channel is a combination of its name along with parameters as follows:

<channel name> ( )

The parameters in the parameter list represent the inputs or outputs of the channel. Each parameter in the list is displayed with the following format:

<direction> : Parameter direction may be in, out, or inout. We formally describe the "channel signature" as a relation $K \subseteq \Lambda \times \Theta$ where $\Lambda$ is a set of "channel names" and $\Theta$ is a set of "parameter lists".

An interaction represents an indivisible and instantaneous communication or handshake between the caller agent (either external environment's actor or component) and the callee agent (component). In the channel-based value-passing approach as shown in Fig. 1, the caller agent interacts with the callee component through the "getPastDueBalance(in studentId: String; out PastDueBalance: Real)" channel signature. Fig. 1 also depicts that the "getPastDueBalance" channel is **required** by the caller and is **provided** by the callee component.

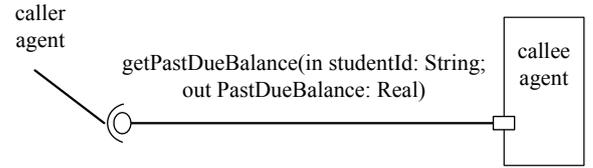

Fig. 1. Channel-based Value-passing Interaction

The external environment uses a "type 1 interaction" to interact with a component. We formally describe the channel-based value-passing "type 1 interaction" as a relation $G \subseteq B \times K \times \Gamma$ where $B$ is a set of "external environment's actors" and $K$ is a set of "channel signatures" and $\Gamma$ is a set of "components".

Two components use a "type 2 interaction" to interact with each other. We formally describe the channel-based value-passing "type 2 interaction" as a relation $V \subseteq \Gamma \times K \times \Gamma$ where $\Gamma$ is a set of "components" and $K$ is a set of "channel signatures".

We can also formally describe the channel-based value-passing "type 1 or 2 interaction" as a relation $\Delta \subseteq \Xi \times K \times \Gamma$ where $\Xi$ is a set of "external environment's actors or components" and $K$ is a set of "channel signatures" and $\Gamma$ is a set of "components".

*B. Entities of SBC Process Algebra*

As shown in Table I, we assume a relation $K$ of channel signatures, and use $k_1, k_2...$ to range over $K$. Further, we let $\Lambda$ be the set of channel names, and use $ch_1, ch_2...$ to range over $\Lambda$. We let $\Theta$ be the set of parameters, and use $p_1, p_2...$ to range over $\Theta$. We let $B$ be the set of actors, and use $\beta_1, \beta_2...$ to range over $B$. We let $\Gamma$ be the set of components, and use $b_1, b_2...$ to range over $\Gamma$. We assume $\Xi$ be the set of actors or components, and use $\rho_1, \rho_2...$ to range over $\Xi$. We let $G$ be the relation of type 1 interactions, and use $g_1, g_2...$ to range over $G$. We let $V$ be the relation of type 2 interactions, and use $v_1, v_2...$ to range over $V$. We let $C$ be a set of guard conditions, and employ $c_1, c_2...$ to range over $C$. Further, we let $\Delta$ be a set of type 1 or 2 interactions, and employ $a_1, a_2...$ to range over $\Delta$. We assume $\Pi$ be the set of code snippets, and employ $\pi_1, \pi_2...$ to range over $\Pi$. We assume $R$ be the set of prefixes, and employ $r_1, r_2...$ to range over $R$. We assume $\Psi$ be the set of state expressions, and employ $s_1, s_2...$ to range over $\Psi$. Further, we assume $\Phi$ be the set of state constants, and employ $A_1, A_2...$ to range over $\Phi$.

TABLE I
ENTITIES OF SBC-PA

| Entity Set | Entity Name | Entity Type |
|---|---|---|
| $K$ | $k_1, k_2...$ | channel signatures |
| $\Lambda$ | $ch_1, ch_2...$ | channel names |
| $\Theta$ | $p_1, p_2...$ | parameter lists |
| $B$ | $\beta_1, \beta_2...$ | actors |
| $\Gamma$ | $b_1, b_2...$ | components |
| $\Xi$ | $\rho_1, \rho_2...$ | actors or components |
| $G$ | $g_1, g_2...$ | type 1 interactions |
| $V$ | $v_1, v_2...$ | type 2 interactions |
| $C$ | $c_1, c_2...$ | guard conditions |
| $\Delta$ | $a_1, a_2...$ | type 1 or 2 interactions |
| $\Pi$ | $\pi_1, \pi_2...$ | code snippets |
| $R$ | $r_1, r_2...$ | prefixes |
| $\Psi$ | $s_1, s_2...$ | states |
| $\Phi$ | $A_1, A_2...$ | state constants |

### C. Prefix of SBC Process Algebra

SBC-PA is a labelled transition system (LTS) which provides a single diagram to unify structural and behavioral constructs. In the SBC-PA transition system, each transition is labelled with a prefix defined as follows.

**DEFINITION** (PREFIX) A prefix $PX = (C, \Delta, \Pi, R)$ consists of

- a finite set $C$ of guard conditions,
- a finite set $\Delta$ of interactions,
- a finite set $\Pi$ of optional code snippets,
- a relation $R \subseteq C \times \Delta \times \Pi$, and $(c, a, \pi) \in R$.

In SBC-PA, all prefixes are guarded by either an explicitly given condition, or the implicit condition [TRUE]. If the value of the condition is TRUE and the interaction is ready, the transition will be triggered. Once the transition is triggered, the appended code snippet will be executed.

### D. Syntax of SBC Process Algebra

As a formal language, SBC-PA is syntactically specified by the Backus-Naur Form (BNF) grammar, as shown in Fig. 2.

(1) $s ::= \bullet$

(2) $s ::= r \bullet s_1$

(3) $s ::= (s_1 \text{ alt } s_2)$

(4) $s ::= (s_1 \text{ par } s_2)$

(5) $s ::= \textbf{loop } ITG$

(6) $s ::= \stackrel{\text{ref}}{=\!=} A$

Fig. 2. BNF grammar of SBC-PA

Rule 1 describes that the inactive state expression "$\bullet$" is a valid SBC-PA state.

Rule 2 describes the state expression "$r \bullet s_1$" will execute the prefix "$r$" first and then continue as the state expression "$s_1$".

Rule 3 describes that the state expression "$s_1$ alt $s_2$" will execute the states "$s_1$" and "$s_2$" alternately.

Rule 4 describes that the state expression "$s_1$ par $s_2$" will interleave the states "$s_1$" and "$s_2$" independently and concurrently.

Rule 5 describes the state, denoted by the state expression "**loop** ITG", a loop definition of infinite behavior represented by the interaction transition graph ITG, is a valid SBC-PA state expression.

Rule 6 describes the state constant "$A$" will be referenced, written as "$\stackrel{\text{ref}}{=\!=} A$".

### E. ITG Overview Diagram

In SBC-PA, each state expression needs to follow the BNF grammar to be grammatically correct. We use the ITG overview diagram (IOD) to represent a grammatically correct state expression.

For example, Fig. 3 shows an ITG overview diagram with the definition:

$s_{34} \stackrel{\text{def}}{=\!=} (s_{35} \text{ par } r_{36} \bullet (\stackrel{\text{ref}}{=\!=} s_{01})) \text{ par } ((\stackrel{\text{ref}}{=\!=} s_{21}) \text{ alt } (\stackrel{\text{ref}}{=\!=} s_{31}))$

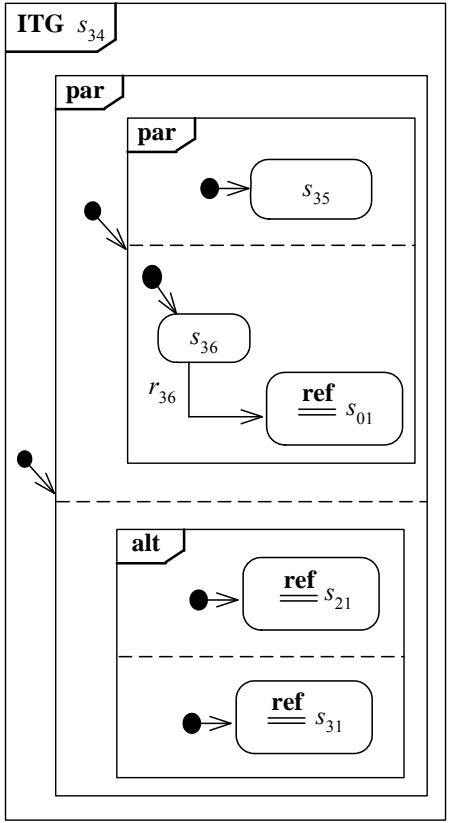

Fig. 3. An example of ITG overview diagram

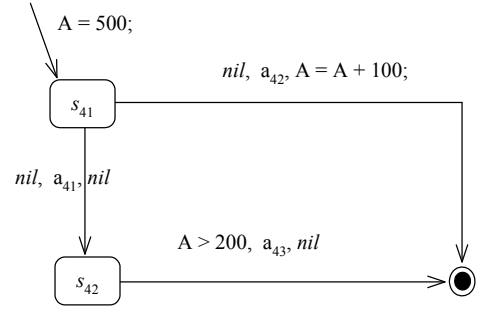

Fig. 4. Interaction transition graph $ITG_{41}$

*F. Transitional Semantics of SBC Process Algebra*

In giving meaning to SBC-PA, we use the interaction transition graph (ITG) as a single diagram to specify the semantics of the system. The interaction transition graph is a labelled transition system (LTS).

**DEFINITION** (INTERACTION TRANSITION GRAPH) An interaction transition graph $ITG = (\Psi, (\pi_0, s_0), R, ITGR)$ consists of

- a finite set $\Psi$ of states,

- an optional code snippet $\pi_0$ in the initial transition, and $\pi_0 \in \Pi$,

- an initial state $s_0 \in \Psi$,

- a relation $R$ of prefixes,

- a transition relation $ITGR \subseteq \Psi_1 \times R \times \Psi_2$, where $(s_j, r, s_k) \in ITGR$ is written as $s_j \xrightarrow{r} s_k$.

For example, the interaction transition relation $ITGR_{41} = \{(s_{41}, (nil, a_{41}, nil), s_{42}), (s_{41}, (nil, a_{42}, A = A + 100;), \bullet), (s_{42}, (A > 200, a_{43}, nil), \bullet)\}$ constitutes the interaction transition graph $ITG_{41} = (\Psi, ((A = 500;), s_{41}), R, ITGR_{41})$, as shown in Fig. 4.

In the interaction transition graph $ITG_{41}$, the state is denoted by a rounded rectangle containing its name; the transition from the source state to the target state is denoted by an arrow and labelled with a guard, an interaction and a code snippet; the initial state (for example, "$s_{41}$") is the target state of the transition that has no source state; the transition without a source state is called the initial transition. In $ITG_{41}$, the code snippet "A = 500;" is attached to the initial transition. In a state, if multiple transitions to be triggered are met, the choice of trigger will be arbitrary and fair.

In the interaction transition graph, whenever $s \xrightarrow{r_1} \cdots \xrightarrow{r_n} s'$, we call $(r_1 \ldots r_n, s')$ a derivative of $s$. For the initial state $s_0$, if $s_0 \xrightarrow{r_1} \cdots \xrightarrow{r_n} s_0$ does occur, then the interaction transition graph is a **loop** definition of the initial state $s_0$. For example, the interaction transition relation $ITGR_{51} = \{(s_{51}, (c\_count > 0, a_{51}, c\_count = c\_count - 1;), s_{52}), (s_{52}, (nil, e_{52}, nil), s_{51}), (s_{51}, (c\_count \leq 0, a_{53}, nil), \bullet)\}$ constitutes the interaction transition graph $ITG_{51} = (\Psi, ((c\_count = 100;), s_{51}), R, ITGR_{51})$, as shown in Fig. 5. In $ITG_{51}$, $s_{51} \xrightarrow{r_{51}} \cdots \xrightarrow{r_{52}} s_{51}$ does occur for the initial state $s_{51}$, therefore $ITG_{51}$ represents the **loop** definition of the state $s_{51}$, that is, $s_{51} = $ **loop** $ITG_{51}$.

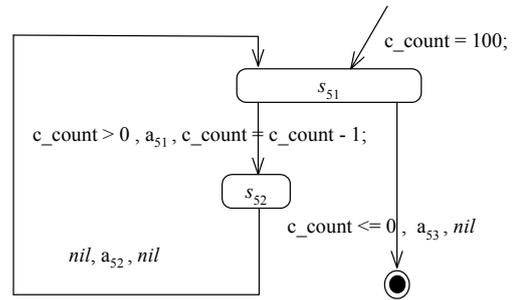

Fig. 5. Interaction transition graph $ITG_{51}$

The semantics for SBC-PA consists in the definition of each *ITGR* which is associated with the transition rules of state operators. These transition rules shall follow the construct of state expressions. Fig. 6 gives the complete set of transition rules: Sequence, Alternative Composition, and Parallel Composition.

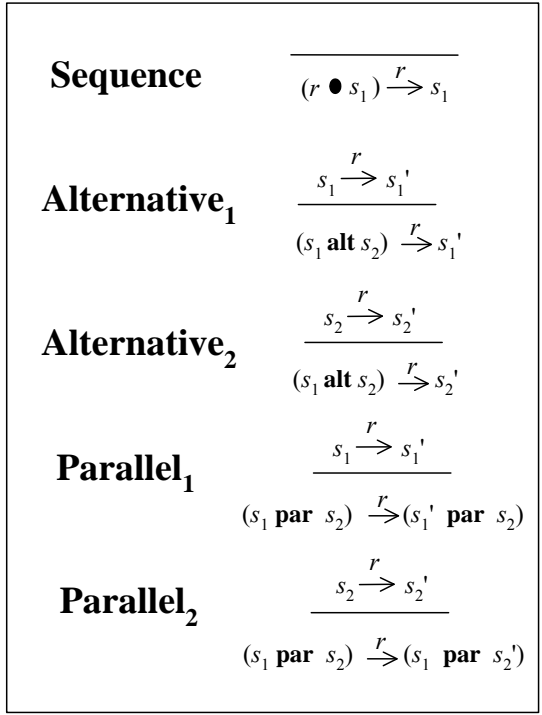

Fig. 6. Transition Rules for SBC-PA

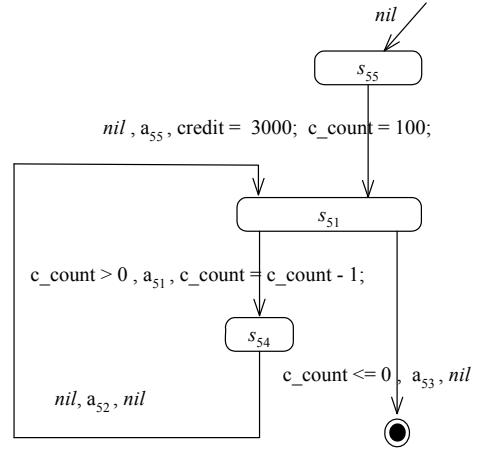

Fig. 7. Interaction transition graph $ITG_{55}$

**Transition Rule of Sequence Composition:** We interpret the Sequence Composition rule as: Under any situation, we generally infer $(r \bullet s_1) \xrightarrow{r} s_1$. To be specific, the state, with a prefix prefixed to it, will use this prefix "$r$" to fulfill the transition.

Let us use "$s_2$" to define the state expression "$r \bullet s_1$", written as "$s_2 \overset{\text{def}}{=\!=} r \bullet s_1$". Sequence Rule updating the initial transition and code snippets on the states "$s_2$" and "$s_1$" is described as follows: (1) The (resulting) code snippet in the prefix "$r$" will be the concatenation of the (original) code snippet in the prefix "$r$" and the code snippet in the initial transition of the state "$s_1$"; (2) The initial transition of the state "$s_1$" will be deleted; (3) The code snippet in initial transition of the state "$s_2$" will be rewritten.

As an example of Sequence Composition transition rule, if we use "$r_{55}$" to define the prefix "$nil, e_{55}$, credit = 3000;" and "$s_{55}$" to define the state expression "$r_{55} \bullet (\overset{\text{ref}}{=\!=} s_{51})$", written as "$s_{55} \overset{\text{def}}{=\!=} r_{55} \bullet (\overset{\text{ref}}{=\!=} s_{51})$", then we will obtain the interaction transition relation $ITGR_{55}$ = {($s_{55}$, ($nil$, $a_{55}$, credit = 3000; c_count = 100;), $s_{51}$), ($s_{51}$, (c_count > 0, $a_{51}$, c_count = c_count - 1;), $s_{54}$), ($s_{54}$, ($nil$, $a_{52}$, $nil$), $s_{51}$), ($s_{51}$, (c_count <= 0, $a_{53}$, $nil$), ●)}, which constitutes the interaction transition graph $ITG_{55}$ = ($\Psi$, ($nil$, $s_{55}$), $R$, $ITGR_{55}$), as shown in Fig. 7.

**Transition Rule of Alternative Composition:** Alternative Composition has two transition rules. Rule Alternative₁ shows that from $s_1 \xrightarrow{r} s_1'$ we will infer $(s_1 \text{ alt } s_2) \xrightarrow{r} s_1'$. Rule Alternative₂ shows that from $s_2 \xrightarrow{r} s_2'$ we shall infer $(s_1 \text{ alt } s_2) \xrightarrow{r} s_2'$.

Let us use "$s_3$" to define the state expression "$s_1 \text{ alt } s_2$", written as "$s_3 \overset{\text{def}}{=\!=} s_1 \text{ alt } s_2$". Alternative Composition updating the initial transition and code snippets on the states "$s_1$" and "$s_2$" and "$s_3$" is described as follows: (1) The code snippet in the state "$s_3$" will be the concatenation of the code snippet in the initial transition of the state "$s_1$" and the code snippet in the initial transition of the state "$s_2$"; (2) The initial transition of the state "$s_1$" will be deleted; (3) The initial transition of the state "$s_2$" will be deleted.

To illustrate the transition rule of Alternative Composition, we construct the interaction transition relation $ITGR_{61}$ = {($s_{61}$, ($nil$, $a_{61}$, $nil$), ●)} which constitutes the interaction transition graph $ITG_{61}$ = ($\Psi$, ($nil$, $s_{61}$), $R$, $ITGR_{61}$), as shown in Fig. 8.

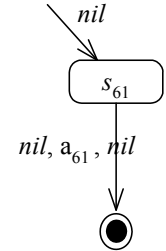

Fig. 8. Interaction transition graph $ITG_{61}$

As an example of Alternative Composition transition rule, if we use "$s_{71}$" to define the state expression "$(\overset{\text{ref}}{=\!=} s_{51}) \text{ alt } (\overset{\text{ref}}{=\!=} s_{61})$", written as "$s_{71} \overset{\text{def}}{=\!=} (\overset{\text{ref}}{=\!=} s_{51}) \text{ alt } (\overset{\text{ref}}{=\!=} s_{61})$", then we will obtain the interaction transition relation $ITGR_{71}$ = {($s_{71}$, (c_count > 0, $a_{51}$, c_count = c_count - 1;), $s_{52}$), ($s_{71}$, (c_count <= 0, $a_{53}$, $nil$), ●), ($s_{71}$, ($nil$, $a_{61}$, $nil$), ●), ($s_{52}$, ($nil$, $a_{52}$, $nil$), $s_{51}$), ($s_{51}$, (c_count > 0, $a_{51}$, c_count = c_count - 1;), $s_{52}$), ($s_{51}$, (c_count <= 0, $a_{53}$, *nil*), ●)} constitutes the interaction transition graph $ITG_{71} = (\Psi, ((c\_count = 100;), s_{71}), R, ITGR_{71})$, as shown in Fig. 9.

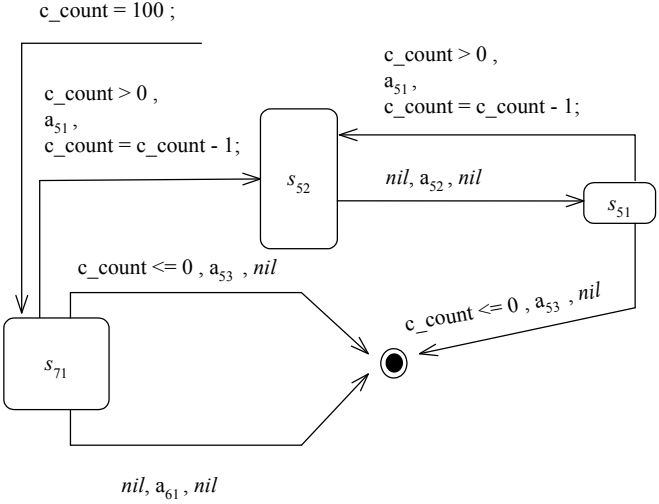

Fig. 9. Interaction transition graph $ITG_{71}$

**Transition Rule of Parallel Composition:** Parallel Composition has two transition rules. Rule Parallel$_1$ shows that from $s_1 \xrightarrow{r} s_1'$ we will infer $(s_1 \text{ par } s_2) \xrightarrow{r} (s_1' \text{ par } s_2)$. Rule Parallel$_2$ shows that from $s_2 \xrightarrow{r} s_2'$ we will infer $(s_1 \text{ par } s_2) \xrightarrow{r} (s_1 \text{ par } s_2')$.

Let us use "$s_4$" to define the state expression "$s_1$ **par** $s_2$", written as "$s_4 \stackrel{\mathbf{def}}{=\!=} s_1$ **par** $s_2$". Parallel Composition updating the initial transition and code snippets on the state "$s_4$" is described as follows: The code snippet in the state "$s_4$" will be the concatenation of the code snippet in the initial transition of the state "$s_1$" and the code snippet in the initial transition of the state "$s_2$".

As an example of Parallel Composition transition rule, if we use "$s_{81}$" to define the state expression "($\stackrel{\mathbf{ref}}{=\!=} s_{51}$) **par** ($\stackrel{\mathbf{ref}}{=\!=} s_{61}$)", written as "$s_{81} \stackrel{\mathbf{def}}{=\!=}$ "($\stackrel{\mathbf{ref}}{=\!=} s_{51}$) **par** ($\stackrel{\mathbf{ref}}{=\!=} s_{61}$)", then we will obtain the interaction transition relation $ITGR_{81} = \{(s_{81},$ (c_count > 0, $e_{51}$, c_count = c_count - 1;), $s_{52}$ **par** $s_{61}$), ($s_{81}$, (c_count <= 0, $a_{53}$, *nil*), ● **par** $s_{61}$), ($s_{81}$, (*nil*, $a_{61}$, *nil*), $s_{51}$ **par** ●), ($s_{52}$ **par** $s_{61}$, (*nil*, $a_{52}$, *nil*), $s_{81}$), ($s_{52}$ **par** $s_{61}$, (*nil*, $a_{61}$, *nil*), $s_{52}$ **par** ●), (● **par** $s_{61}$, (*nil*, $a_{61}$, *nil*), ● **par** ●), ($s_{52}$ **par** ●, (*nil*, $a_{52}$, *nil*), $s_{51}$ **par** ●), ($s_{51}$ **par** ●, (c_count <= 0, $a_{53}$, *nil*), ● **par** ●), ($s_{51}$ **par** ●, (c_count > 0, $a_{51}$, c_count = c_count - 1;), $s_{52}$ **par** ●)} constitutes the interaction transition graph $ITG_{81} = (\Psi, ((c\_count = 100;), s_{81}), R, ITGR_{81})$, as shown in Fig. 10.

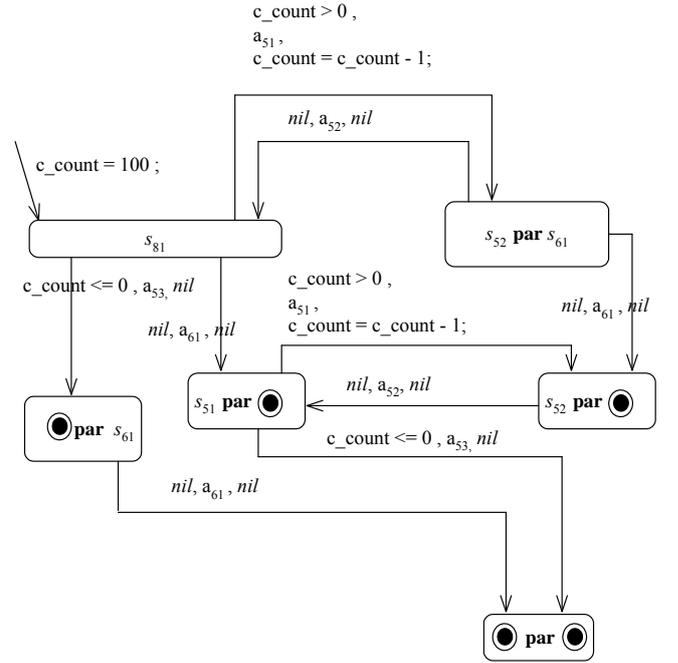

Fig. 10. Interaction transition graph $ITG_{83}$

The complete state space in the parallel composition is the Cartesian product of the two states. Although it is correct, it is not readable. Therefore, sometimes we will express the interaction transition graph without executing the parallel composition transition rule.

For example, the interaction transition graph "$ITG_{81}$" shown in Fig. 10 defining the state expression "($\stackrel{\mathbf{ref}}{=\!=} s_{51}$) **par** ($\stackrel{\mathbf{ref}}{=\!=} s_{61}$)", does execute the parallel composition transition rule. Contrary to "$ITG_{81}$", the interaction transition graph "$ITG_{91}$" defining the state expression "($\stackrel{\mathbf{ref}}{=\!=} s_{51}$) **par** ($\stackrel{\mathbf{ref}}{=\!=} s_{61}$)", does not execute the parallel composition transition rule, as shown in Fig. 11.

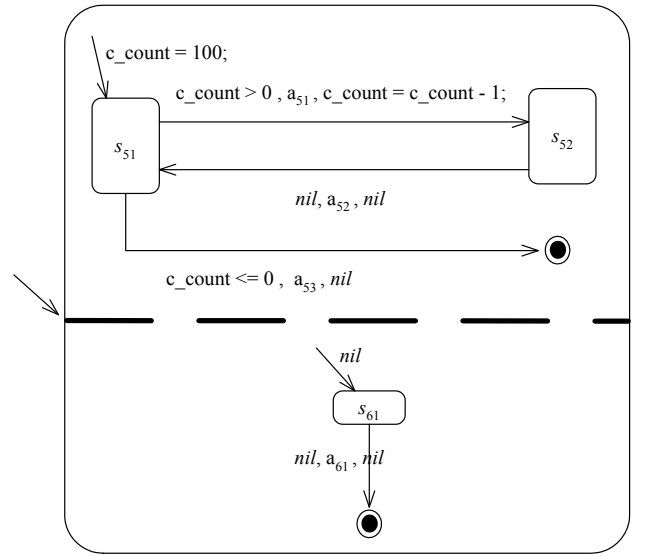

Fig. 11. Interaction transition graph $ITG_{91}$

Although the appearances of the interaction transition diagrams "$ITG_{81}$" and "$ITG_{91}$" look so different, they are equivalent in behavior.

### IV. CASE STUDY: AUTOMATED TELLER MACHINE

The automated teller machine (ATM) offers withdrawal of cash if the bank card ID and PIN code are successfully verified, and has cash refill such that the customers can get what they order.

An ATM operator may shut down the ATM for routine maintenance. Fig. 12 shows the ITG overview diagram of ATM with the following definition:

$$s_{\text{ATM}} \stackrel{\text{def}}{=\!=} (\stackrel{\text{ref}}{=\!=} s_{101}) \text{ par } (\stackrel{\text{ref}}{=\!=} s_{201}) \text{ par } (\stackrel{\text{ref}}{=\!=} s_{301})$$

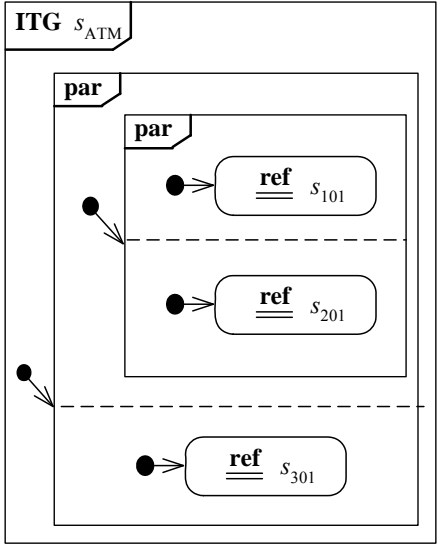

Fig. 12. IOD of $ITG_{\text{ATM}}$

The interaction transition relation $ITGR_{101}$ = {($s_{101}$, (nil, (Customer, inputCardInformation(in cardId; in PIN), :ATM), nil), $s_{102}$), ($s_{102}$, (nil, (:ATM, validatePIN(in cardId; in PIN; out cardValid; out accountId), :Bank), nil), $s_{103}$), ($s_{103}$, (cardValid = "yes", (Customer, withdrawCash(in amount), :ATM), nil), $s_{104}$), ($s_{104}$, (nil, (:ATM, retrieveBalance(in accountId; out balance), :Bank), nil), $s_{105}$), ($s_{105}$, (balance > amount, (Customer, dispenseCash(out cash), :ATM), nil), $s_{101}$)}. Fig. 13 shows the interaction transition graph $ITG_{101}$ which is represented by $s_{101}$.

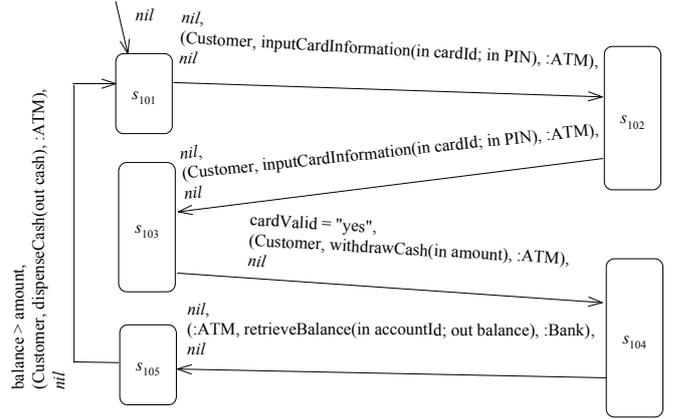

Fig. 13. $ITG_{101}$

The interaction transition relation $ITGR_{201}$ = {($s_{201}$, (nil, (Operator, refillCash(in cash), :ATM), nil), $s_{201}$)} constitutes the interaction transition graph $ITG_{201}$ = ($\Psi$, (nil, $s_{201}$), R, $ITGR_{201}$). Fig. 14 shows the interaction transition graph $ITG_{201}$ which is represented by $s_{201}$.

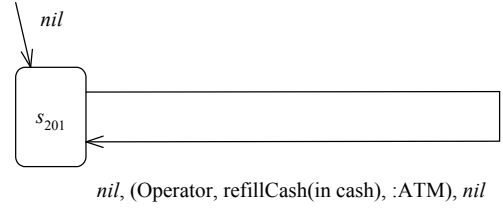

nil, (Operator, refillCash(in cash), :ATM), nil

Fig. 14. $ITG_{201}$

The interaction transition relation $ITGR_{301}$ = {($s_{301}$, (nil, (Operator, shutDown, :ATM), nil), $s_{301}$)} constitutes the interaction transition graph $ITG_{301}$ = ($\Psi$, (nil, $s_{301}$), R, $ITGR_{301}$). Fig. 15 shows the interaction transition graph $ITG_{301}$ which is represented by $s_{301}$.

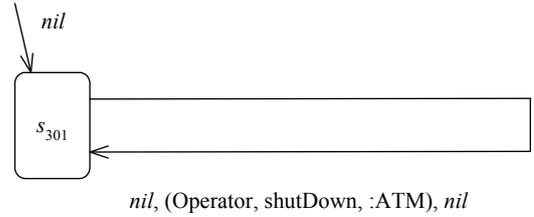

nil, (Operator, shutDown, :ATM), nil

Fig. 15. $ITG_{301}$

### V. CONCLUSIONS

In the 1920s, Ludwig von Bertalanffy wrote: there exist models, principles, and laws that apply to generalized systems or their subclasses, irrespective of their particular kind, the nature of their constituent elements, and the relationships or "forces" between them [13]. In this paper, we refer what Bertalanffy proposed and developed as the systems definition 1.0. Systems definition 1.0 describes a system hopefully to be an integrated whole, embodied in its components, their interrelationships with each other and the environment, and the principles and guidelines governing its design and evolution. This systems definition 1.0 possesses one cardinal deficiency. The deficiency comes from that it does not define the integration of systems structure and systems behavior. Systems structure and systems behavior are the two most significant views of a system. In order to achieve a truly integrated whole of a system, we first need to integrate the systems structure and behavior together. In other words, integration of the systems structure and systems behavior results in the integration of a whole system. Since systems definition 1.0 does not describe the integration of systems structure and systems behavior, very likely it only hopes and will never be able to actually form an integrated whole of a system. In this situation, systems definition 1.0 is powerless in defining a system suitably.

Structure-behavior coalescence (SBC) architecture provides an elegant way to integrate the structure and behavior of a system. A system is therefore redefined, by systems definition 2.0, truly to be an integrated whole, using the SBC architecture, embodied in its assembled components, their interactions with each other and the environment, and the principles and guidelines governing its design and evolution. Since systems definition 2.0 describes the integration of systems structure and systems behavior, definitely it is able to form an integrated whole of a system. In this situation, systems definition 2.0 is fully capable of defining the system.

ACKNOWLEDGEMENTS

The author is thankful to the anonymous reviewers for their useful remarks, which help illuminate the nuances and sparked new ideas.

**William S. Chao** was born in 1954 in Taiwan and received his Ph.D. degree in information science from the University of Alabama at Birmingham, USA, in 1988. William worked as a computer scientist at GE Research and Development Center, USA, from 1988 till 1991 and has been teaching at National Sun Yat-Sen University, Taiwan from 1992 till 2019. His research covers: systems architecture, hardware architecture, software architecture, and enterprise architecture. Dr. Chao is a member of the Chinese Association of Enterprise Architects.